\documentclass[sigconf,nonacm]{acmart}
\usepackage{amsmath}
\usepackage{algorithm}
\usepackage{algpseudocode}
\AtBeginDocument{%
  }

\setcopyright{none}
\copyrightyear{2026}
\acmYear{2026}
\acmDOI{XXXXXXX.XXXXXXX}

\renewcommand\footnotetextcopyrightpermission[1]{}

\begin{document}

\title{KPage: Cross-Block Page Compaction for Hardware-Efficient Sparse Attention}


\settopmatter{authorsperrow=3}

\author{Kling Infra Team}
\affiliation{%
  \institution{KlingAI Research}
    \city{Beijing}
  \country{China}
}

\makeatletter
\g@addto@macro\addresses{%
  \and\par\noindent\hspace*{-1em}\let\and\relax
}
\makeatother

\author{Junwei Feng}
\authornote{These authors contributed equally to this work.}
\authornote{This work was conducted during the author's internship at KlingAI.}
\email{fengjunwei@kuaishou.com}
\affiliation{%
  \institution{KlingAI Research}
  \city{Beijing}
  \country{China}
}

\author{Qiaorui Chen}
\authornotemark[1]
\email{cjerry@nvidia.com}
\affiliation{%
  \institution{NVIDIA}
  \city{Shanghai}
  \country{China}
}

\author{Chenyu Liang}
\email{liangchenyu03@kuaishou.com}
\affiliation{%
  \institution{KlingAI Research}
  \city{Beijing}
  \country{China}
}

\author{Zhen Guo}
\email{jemiryg@nvidia.com}
\affiliation{%
  \institution{NVIDIA}
  \city{Shanghai}
  \country{China}
}

\author{Jingyi Yang}
\email{sopyang@nvidia.com}
\affiliation{%
  \institution{NVIDIA}
  \city{Beijing}
  \country{China}
}

\author{Xiaoling Zhou}
\email{zhouxiaoling03@kuaishou.com}
\affiliation{%
  \institution{KlingAI Research}
  \city{Beijing}
  \country{China}
}

\author{Jiahao Zhang}
\email{zhangjiahao09@kuaishou.com}
\affiliation{%
  \institution{KlingAI Research}
  \city{Beijing}
  \country{China}
}

\author{Dongliang Huang}
\email{huangdongliang@kuaishou.com}
\affiliation{%
  \institution{KlingAI Research}
  \city{Beijing}
  \country{China}
}

\author{Yiqiao Liao}
\authornotemark[3]
\email{liaoyiqiao@kling.ai}
\affiliation{%
  \institution{KlingAI Research}
  \city{Beijing}
  \country{China}
}

\author{Chengru Song}
\authornote{Corresponding authors.}
\email{songchengru@kuaishou.com}
\affiliation{%
  \institution{KlingAI Research}
  \city{Beijing}
  \country{China}
}


\begin{abstract}
Sparse attention reduces the algorithmic cost of long-sequence video
Transformers, but three-dimensional token layouts can leave selected
block-sparse regions only partially occupied. As a result, eliminating
unselected blocks does not necessarily eliminate unused capacity inside
the physical tiles that remain.

We present KPage, a page-compacted execution scheme for block-sparse
attention. KPage decomposes selected K/V blocks into 16-row pages and
packs valid pages across logical-block boundaries into large physical
attention tiles. The packed tiles are assembled directly during K/V
loading, without materializing an intermediate compacted tensor in
high-bandwidth memory.
KPage further routes partially occupied query blocks between
$M64\times N256$ and $M128\times N128$ forward kernels, and uses exact
score masking to preserve the original sparse-attention semantics.
For backward, KPage constructs a K/V-owned paged layout and restores
the original sparse adjacency after packing both Q and K/V.

Across five representative sparse-attention workload cases, KPage achieves
$1.35\times$--$2.06\times$ bare forward-kernel speedup over NVIDIA
Block-Sparse Attention (BSA),
with a geometric mean of $1.77\times$. On controlled workloads with
fully occupied logical blocks, KPage is effectively on par with BSA
($1.03\times$ geometric mean). Under matched-scale 3D layouts containing
partial blocks, the geometric-mean speedup increases to $2.29\times$.
Execution-capacity analysis shows that Page16 K/V compaction and
query-adaptive execution reduce physical query--key pair capacity from
334.1G to 177.5G positions while preserving 160.6G useful interactions.
Across eight layouts reconstructed from the open-source Selective and
Sliding Tile Attention (SSTA), Sliding Tile Attention (STA), and Video
Sparse Attention (VSA) systems, the same KPage kernel has the lowest forward time in every
case: 1.13--4.25$\times$ faster than the strongest exact baseline.
The corresponding forward-plus-backward ratios are
1.02--2.71$\times$.  We further extend the metadata interface to
independent Q/K partitions and evaluate the paper-matched Sparse
VideoGen2 (SVG2) Wan 14B/720p configuration.  KPage is $1.49\times$ faster than the
strongest exact sparse baseline and $1.57\times$ faster than NVIDIA BSA. After compliance review, we will open-source KPage and submit a PR to NVIDIA cuDNN Frontend.
\end{abstract}


\maketitle
\pagestyle{plain}
\thispagestyle{plain}


\section{Introduction}

Self-attention is the core computation in Transformer
architectures~\cite{vaswani2017attention}, and its quadratic complexity
becomes particularly expensive in modern video diffusion Transformers
(DiTs)~\cite{peebles2023dit}. State-of-the-art video generators
typically operate on compressed spatiotemporal representations and use
Transformer backbones over long token sequences
~\cite{blattmann2023stablevideo,yang2025cogvideox,
kong2024hunyuanvideo,wan2025}. In particular, modern video variational
autoencoders (VAEs) compress video along both temporal and spatial dimensions, producing
an underlying three-dimensional token organization that can be written
as $T\times H\times W$. Once flattened for Transformer execution,
these representations can contain tens or hundreds of thousands of
tokens, making attention one of the dominant costs of both training
and inference.

Sparse attention is a natural way to reduce this cost. Earlier work
established structured sparse attention for long sequences using
strided, local, global, or mixed connectivity
~\cite{child2019sparsetransformer,beltagy2020longformer,
zaheer2020bigbird}. For video generation, the spatial and temporal
structure provides substantially stronger sparsity opportunities.
Recent systems exploit either structured spatiotemporal locality or
content-dependent sparse selection. Sparse VideoGen identifies
spatial- and temporal-dominant attention heads
~\cite{xi2025sparsevideogen}; STA converts
3D local attention into hardware-friendly tiles~\cite{zhang2025sta};
VSA learns dynamic 3D block selection
~\cite{zhang2025vsa}; SVG2 uses semantic clustering and
permutation to densify selected regions
~\cite{yang2025sparsevideogen2}; and Radial Attention constructs a
distance-dependent sparse pattern for long videos
~\cite{li2025radial}. SSTA further combines selective and sliding-tile
attention in HunyuanVideo 1.5~\cite{wu2025hunyuanvideo15}. Other recent
systems, including SpargeAttention, VORTA, FPSAttention, and SPADE,
likewise exploit structured or dynamic sparsity to reduce video
attention cost
~\cite{zhang2025spargeattn,sun2025vorta,
liu2025fpsattention,liu2026spade}.

Despite their different sparse-selection mechanisms, efficient
implementations of these methods share an important property: sparse
attention is eventually quantized into block- or tile-level GPU
execution. This is necessary because reducing arithmetic work alone
does not guarantee wall-clock speedup. Efficient attention kernels
depend on sufficiently large matrix operations, regular memory
movement, and fused online softmax. Consequently, sparse selection can identify useful interactions at a
finer granularity than the tiles a GPU executes efficiently.

\begin{table*}[t]
\centering
\caption{
Forward bare-kernel performance on padding-free and 3D-padded VAE
layouts. Length is the physical sequence length after layout
construction, and sparsity is measured over valid token pairs.
Within each row, BSA and KPage execute the same sparse graph and
valid-row metadata. Speedup is BSA/KPage; displayed latencies and
reported ratios are rounded independently.
}
\label{tab:padding_ablation}

\small
\setlength{\tabcolsep}{6pt}
\renewcommand{\arraystretch}{1.12}

\begin{tabular}{c r r r r r r r r r r r}
\toprule
&
\multicolumn{5}{c}{\textbf{Padding-free (0\% padding)}} &
\multicolumn{6}{c}{\textbf{3D padded}} \\
\cmidrule(lr){2-6}
\cmidrule(lr){7-12}

\textbf{Video}
& \textbf{Length}
& \textbf{Sparsity}
& \textbf{BSA}
& \textbf{KPage}
& \textbf{Speedup}
& \textbf{Length}
& \textbf{Sparsity}
& \textbf{Padding}
& \textbf{BSA}
& \textbf{KPage}
& \textbf{Speedup} \\

&
&
&
\multicolumn{2}{c}{\textbf{Time (ms)}}
&
&
&
&
&
\multicolumn{2}{c}{\textbf{Time (ms)}}
& \\
\midrule

5s
& 65,536 & 87.33\%
& 1.65 & \textbf{1.50} & \textbf{1.11$\times$}
& 81,920 & 92.68\% & 22.8\%
& 1.86 & \textbf{1.08} & \textbf{1.73$\times$} \\

10s
& 122,880 & 93.24\%
& 3.38 & \textbf{3.28} & \textbf{1.03$\times$}
& 158,720 & 97.00\% & 21.6\%
& 3.82 & \textbf{1.74} & \textbf{2.20$\times$} \\

15s
& 184,320 & 95.49\%
& 5.32 & \textbf{5.03} & \textbf{1.06$\times$}
& 235,520 & 98.29\% & 21.2\%
& 5.66 & \textbf{2.32} & \textbf{2.44$\times$} \\

30s
& 368,640 & 97.74\%
& 12.17 & \textbf{12.97} & \textbf{0.94$\times$}
& 465,920 & 99.34\% & 20.8\%
& 11.70 & \textbf{3.97} & \textbf{2.95$\times$} \\

\midrule
\textbf{Geometric Mean}
& -- & --
& -- & -- & \textbf{1.03$\times$}
& -- & -- & --
& -- & -- & \textbf{2.29$\times$} \\

\bottomrule
\end{tabular}
\end{table*}

This mismatch is particularly important for three-dimensional
video layouts. A logical sparse block is defined over a flattened token
sequence, while token validity is inherited from the original
$T\times H\times W$ geometry. If a temporal or spatial dimension is
not aligned with the blocking geometry, boundary blocks contain fewer
valid tokens than their nominal block size. Unlike one-dimensional
sequence padding, which is commonly concentrated at the end of a
sequence, multidimensional boundaries recur along temporal and spatial
surfaces and their intersections. After flattening, partially occupied
blocks can therefore appear repeatedly throughout the sequence.

This is not a hypothetical corner case. Existing video sparse-attention
systems have independently encountered closely related forms of
intra-block under-utilization. STA identifies \emph{mixed blocks}, in
which only part of a hardware tile belongs to the desired 2D or 3D
attention window, and redesigns the sliding pattern so that executed
tiles are predominantly fully dense or fully empty
~\cite{zhang2025sta}. SVG2 observes the same granularity
problem for dynamically sized semantic clusters: a useful
$128\times32$ attention region executed with a fixed
$128\times128$ sparse kernel requires 75\% of the physical computation
to be spent outside the useful region
~\cite{yang2025sparsevideogen2}. VSA exposes the complementary
hardware trade-off: smaller sparse blocks provide finer selection
granularity, but reducing the block size lowers kernel throughput;
its implementation therefore uses a $64\times64$ block-sparse kernel
and carefully balances sparsity against hardware efficiency
~\cite{zhang2025vsa}.

The difficulty is that simply shrinking the compute tile is not free.
The evolution from FlashAttention to FlashAttention-2 (FA2),
FlashAttention-3 (FA3), and FA4 has progressively increased the
importance of hardware-aware tiling, asynchronous data movement, and
deeply pipelined matrix operations
~\cite{dao2022flashattention,dao2024flashattention2,
shah2024flashattention3,zadouri2026flashattention4}.
ThunderKittens similarly builds high-performance attention around
regular tile abstractions and asynchronous pipelines
~\cite{spector2024thunderkittens}, while FlashInfer and FlexAttention
retain block-structured execution to combine flexibility with high
GPU efficiency~\cite{ye2025flashinfer,dong2025flexattention}.
Thus, smaller execution granularity can remove padding but can also
give up the efficiency of large attention tiles.

Current systems address this tension in several ways. STA changes the
sparse pattern so that mixed blocks disappear. SVG2
changes the compute geometry itself by supporting dynamically sized
blocks. General block-sparse executors such as NVIDIA BSA retain a
fixed block representation while accepting valid-length metadata so
that padded K/V rows are excluded from the mathematical attention
result~\cite{nvidia2026bsa}. For fixed-tile block-sparse execution, valid rows from different
selected logical blocks remain uncombined, leaving physical tile
capacity underutilized. Fine-grained sparse execution such as
FG-Attn gathers selected K/V vectors into packed shared-memory tiles
~\cite{durvasula2025fgattn}, but its fine-grained units represent the
attention sparsity pattern itself rather than invalid rows inside an
already selected 3D sparse block.

We present KPage, which removes partial-block waste by packing data
rather than shrinking the large compute tile. KPage keeps the original
blk128 sparse graph but divides selected K/V blocks into 16-row pages.
Valid Page16 regions from different logical blocks are compacted across
block boundaries and gathered directly into the shared-memory stages
of the attention kernel. No compacted K/V tensor is materialized in
HBM. The resulting physical tile remains large enough for efficient
Tensor Core execution while containing substantially less padding.

KPage also addresses partial occupancy on the query dimension. Query
blocks with at most 64 valid rows use an $M64\times N256$ path, while
larger blocks use $M128\times N128$; both geometries contain the same
16,384 query--key positions per physical tile. Invalid rows in partial
pages are masked before the online-softmax row maximum, preserving
exact attention semantics. For backward, KPage constructs a K/V-owned
paged layout, derives reverse-Q neighborhoods for packed K/V tiles, and
uses page-pair adjacency metadata to restore the original Q--K sparse
graph after both sides have been physically packed.

The page abstraction does not require Q and K/V to share one logical
partition.  We expose separate per-head Q block sizes and K/V block
sizes and offsets, allowing independently clustered layouts such as
SVG2 to reuse the same page execution.  Q clusters retain the
Q64/Q128 routes, while independently sized K/V clusters are placed on
Page16 boundaries.

Our evaluation shows that the gain is specifically associated with
partial-block fragmentation rather than a generic advantage over the
block-sparse baseline. When every logical block is fully occupied,
KPage and NVIDIA BSA are effectively on par, with a
$1.03\times$ geometric-mean forward speedup. Under matched-scale
workloads containing 3D partial blocks, KPage achieves a
$2.29\times$ geometric-mean forward speedup. Across five representative
sparse-attention workload cases, KPage achieves a $1.77\times$
geometric-mean bare forward-kernel speedup over BSA. Page16 K/V
compaction and query-adaptive execution reduce physical query--key pair
capacity from 334.1G to 177.5G positions while preserving 160.6G useful
interactions, increasing pair-capacity efficiency from 48.1\% to
90.5\%.  On the paper-matched SVG2 Wan 14B/720p configuration, the
independent-partition extension is $1.49\times$ faster than the
strongest exact sparse baseline.

\subsection{Summary of Contributions}
Our contributions are as follows:
\begin{itemize}
\item We propose KPage, which packs valid Page16 regions across logical K/V block boundaries directly into large attention tiles, without materializing an intermediate compacted tensor in HBM. Query-adaptive tiling reduces Q-side padding, while K/V-owned page packing preserves the sparse graph in backward execution.

\item We extend the same execution scheme to independent Q/K partitions, supporting sparse layouts with different query and key/value block boundaries, including semantic clustering.

\item Across five representative workloads, KPage achieves a \textbf{$1.77\times$} geometric-mean forward speedup over NVIDIA BSA, and \textbf{$2.29\times$} on controlled 3D partial-block layouts. Across eight open-source SSTA, STA, and VSA layouts, KPage is \textbf{$1.13$--$4.25\times$} faster than the strongest exact sparse forward baseline. It further achieves \textbf{$1.49\times$} on the SVG2 Wan 14B/720p configuration.
\end{itemize}

\begin{figure*}[htbp]
    \centering
    \includegraphics[width=\textwidth]{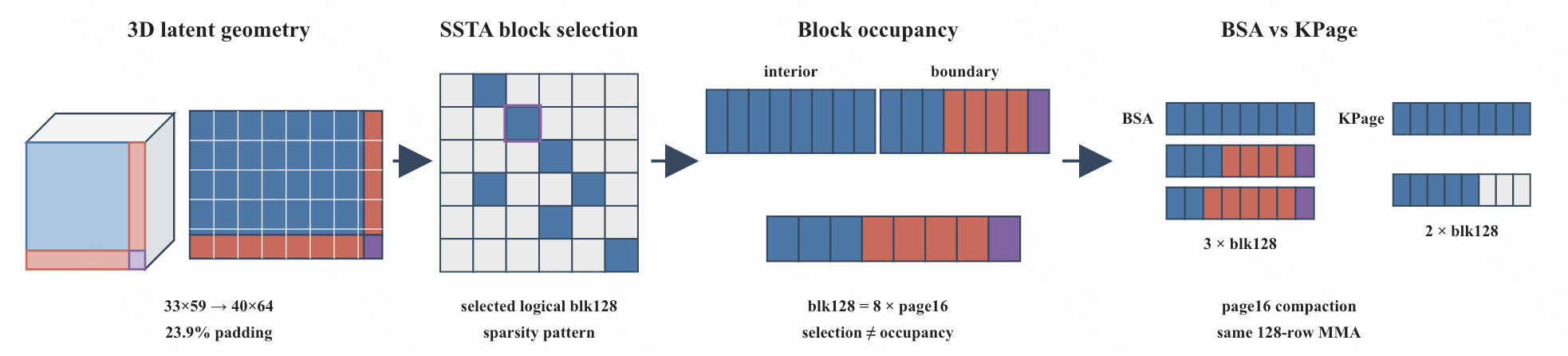}
\caption{
Three-dimensional video layouts can produce partially occupied
logical blocks after flattening. A block-sparse selector may retain
such blocks even though only a subset of their rows are valid.
KPage decomposes selected blocks into Page16 regions and compacts
valid pages across logical-block boundaries while retaining large
Tensor Core execution tiles.
}
    \label{fig:intro}
\end{figure*}

\section{State of the Art}
\label{sec:sota}

Recent video sparse-attention systems have made substantial progress in
both sparse-pattern design and GPU execution. Sparse VideoGen exploits
spatial--temporal sparsity in pretrained video diffusion models and
combines the sparse pattern with layout transformation and customized
sparse kernels~\cite{xi2025sparsevideogen}. STA targets local 3D
attention and redesigns token-wise sliding windows into
hardware-aligned tiles; its ThunderKittens-based implementation reaches
58.79\% model FLOPs utilization (MFU) and accelerates attention by
$1.6$--$10\times$ over FA3 depending on the workload
~\cite{zhang2025sta,spector2024thunderkittens,
shah2024flashattention3}. VSA performs dynamic selection over 3D video
blocks and implements dedicated block-sparse forward and backward
kernels, reaching more than 85\% of the MFU of FA3 while
substantially reducing attention time at high sparsity
~\cite{zhang2025vsa}. These results demonstrate that modern video
sparse attention is already heavily co-designed with high-performance
GPU kernels.

Other systems follow similar hardware-aware designs. Radial Attention
constructs structured spatiotemporal sparse masks and executes them
with FlashInfer for inference and an FA2-derived
block-sparse kernel for training
~\cite{li2025radial,ye2025flashinfer,
mit2024blocksparse,dao2024flashattention2}. HunyuanVideo 1.5 deploys
SSTA through a dedicated flex-block-attention library built on
ThunderKittens; the implementation supports STA, selective sparse
attention, and their hybrid for both training and inference
~\cite{wu2025hunyuanvideo15,spector2024thunderkittens}.
SVG2 goes further by supporting dynamically sized sparse
regions. It observes that semantic clustering can produce, for example,
a $128\times32$ useful region that would waste 75\% of the computation
if executed as a fixed $128\times128$ block, and develops
dynamic-block-size attention kernels following FlashAttention-style
execution~\cite{yang2025sparsevideogen2}. SVG2 reports up to
$2.30\times$ and $1.89\times$ end-to-end speedup for HunyuanVideo and
Wan on H100, respectively.  Our SVG2 study in
Section~\ref{sec:eval_generality} instead isolates attention-kernel
execution and is not a reproduction of those end-to-end results.
Collectively, these systems
show that the state of the art already includes sophisticated sparse
layouts, specialized block-sparse kernels, asynchronous data movement,
and variable execution geometries.

NVIDIA BSA represents the corresponding state of the art in general
block-sparse execution~\cite{nvidia2026bsa}. Its CUTLASS DSL kernels
consume per-query lists of selected K/V blocks and support explicit
valid lengths for partially occupied K/V blocks in forward. Thus,
partial blocks can already be represented with exact mathematical
semantics rather than treating padding as valid attention input.
BSA nevertheless preserves the selected logical block as the unit of
the sparse execution interface: valid rows from different logical
blocks are not physically combined into the same K/V tile. The current
blk128 backward path additionally requires full physical K/V blocks.

The above systems optimize different sources of sparse-attention
inefficiency, but they do not target the particular padding studied in
this work. STA removes \emph{mixed attention tiles} created by the
sliding-window mask by changing the sparse pattern itself. SVG2
removes padding caused by \emph{variable semantic-cluster
sizes} by varying the compute-block geometry. VSA studies the trade-off
between sparse block size and kernel throughput. In contrast, our
problem appears \emph{after} a logical sparse block has already been
selected: because valid tokens originate from a three-dimensional
video layout, a selected blk128 block can contain substantially fewer
than 128 valid rows. This layout-induced partial occupancy is
independent of which sparse-selection algorithm produced the block.

Existing block-sparse kernels therefore leave a specific execution
opportunity. A partially occupied selected block need not be executed
as an independent under-filled physical tile if its valid subregions
can be combined with valid subregions from other selected blocks.
To the best of our knowledge, prior video sparse-attention kernels do
not perform such cross-logical-block packing of valid sub-blocks while
retaining a large physical attention tile.

KPage targets exactly this remaining inefficiency. It decomposes the
valid K/V region of each selected blk128 block into Page16 units and
packs pages from different logical blocks into the same physical K/V
tile. The packed tile is assembled directly during K/V loading, without
materializing a compacted K/V tensor in HBM. As a result, KPage retains
the large Tensor Core execution geometry used by high-performance
attention kernels while substantially increasing the occupancy of each
physical tile. The sparse graph and its selection policy remain
unchanged.

\section{Methodology and Implementation}
\label{sec:method}

\subsection{Overview}
\label{sec:overview}
KPage retains the logical blk128 sparse graph while changing its
mapping to physical attention tiles. In forward, selected K/V blocks
are divided into 16-row pages and valid pages from different blocks
are packed into the same large tile. Two query tile shapes reduce
Q-side padding without merging Q blocks with different sparse
neighborhoods.

Backward requires a different organization: multiple Q blocks may
contribute to the gradients of the same K/V block. KPage therefore
packs pages into K/V-owned tiles, traverses their reverse-Q
neighborhoods, and records page-pair adjacency to preserve the
original sparse graph. Algorithm~\ref{alg:kpage} summarizes the
execution flow.

\subsection{Page-Granular Forward Execution}
\label{sec:forward}

\subsubsection{K/V page compaction}

For logical K/V block $b$, let $c_b\in[0,B]$ be its valid-row
count and $p_b$ its number of Page16 regions:

\begin{equation}
p_b=\left\lceil\frac{c_b}{P}\right\rceil,
\qquad
0\le Pp_b-c_b<P ,
\label{eq:num_pages}
\end{equation}

where $P=16$. Thus, the physical extent of a selected block decreases
from 128 rows to $16p_b$ rows before final tile alignment. The
remaining padding within each logical block is bounded by 15 rows.

For a query block $q$, let $\mathcal{K}_q$ be the ordered list of
logical K/V blocks selected by the sparse-attention algorithm. KPage
expands these blocks into Page16 regions. Fully valid pages from all
selected blocks are emitted first, followed by at most one partial tail
page from each partially occupied block. Dummy pages are appended only
when required by the physical iteration granularity.

This ordering serves two purposes. First, pages from different logical
blocks can share a physical K/V tile, eliminating the requirement that
each selected blk128 begin at a separate tile boundary. Second, the
partial pages are confined to a suffix of the page walk. The kernel can
therefore identify a prefix consisting entirely of valid rows and
execute it without loading validity metadata.

The sparse graph is not changed by this transformation. A page is
included only if its parent logical K/V block belongs to
$\mathcal{K}_q$; KPage changes the physical packing of selected blocks,
not the set of selected blocks.

\begin{figure}[htbp]
    \centering
    \includegraphics[width=\linewidth]{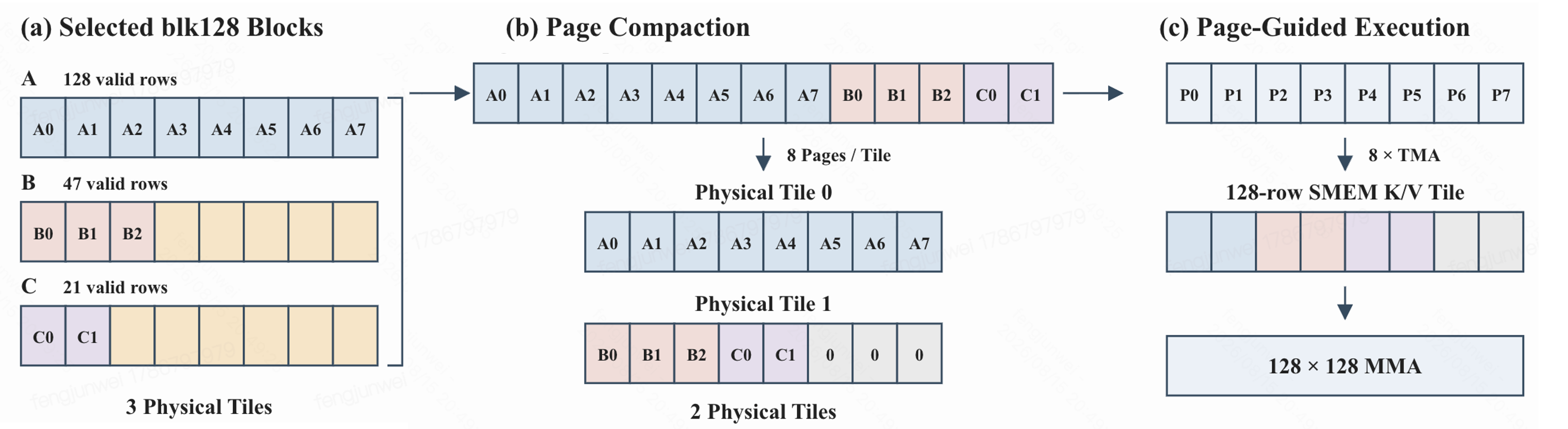}
    \caption{
    K/V page compaction. Selected blk128 blocks are decomposed into
    Page16 regions. Fully valid pages are packed across logical-block
    boundaries, followed by partial tail pages. Eight Page16 regions
    form one physical $N128$ K/V tile.
    }
    \label{fig:page-layout}
\end{figure}

\subsubsection{Query-adaptive tile shapes}

Page compaction reduces padding on K/V but does not reduce padding
inside the Q block. Packing several Q blocks together would require
merging their K/V neighborhoods, because different Q blocks generally
have different sparse selections. KPage instead keeps every logical
Q block independent and selects the physical tile shape from its
valid-row count $c_q$, where $M_q$ and $N_q$ are the Q and K/V tile
dimensions:

\begin{equation}
(M_q,N_q)=
\begin{cases}
(64,256),  & 0<c_q\le64,\\
(128,128), & 64<c_q\le128.
\end{cases}
\label{eq:q_route}
\end{equation}

Empty Q blocks are skipped.

Both paths expose the same number of Q--K positions per physical tile:
$64\times256=128\times128=16{,}384$. The Q64 path therefore reduces
unused Q rows without reducing the tile area presented to Tensor Core.
It consumes sixteen Page16 K/V regions per iteration; the Q128 path
consumes eight.

The two optimizations operate independently. Page compaction reduces
waste in the K/V dimension, while Q64/Q128 routing reduces waste in the
Q dimension. Neither operation changes the sparse neighborhood of the
query block.

Query blocks from the two paths are placed in separate persistent work
queues. Within each route, longer page walks are scheduled earlier to
reduce the tail caused by uneven sparse workloads. The underlying
pipeline requires aligned iteration counts, so a small number of dummy
pages can remain at the end of a walk. These pages are handled by the
same validity mechanism as partial pages.

\subsubsection{Direct page-guided K/V loading}

Materializing the compacted K/V layout in HBM would require an additional gather kernel followed by another HBM read by the attention kernel. Instead, KPage uses the page table to assemble packed K/V tiles directly in shared memory during data loading, avoiding intermediate HBM materialization.

\begin{figure}[htbp]
    \centering
    \includegraphics[width=\linewidth]{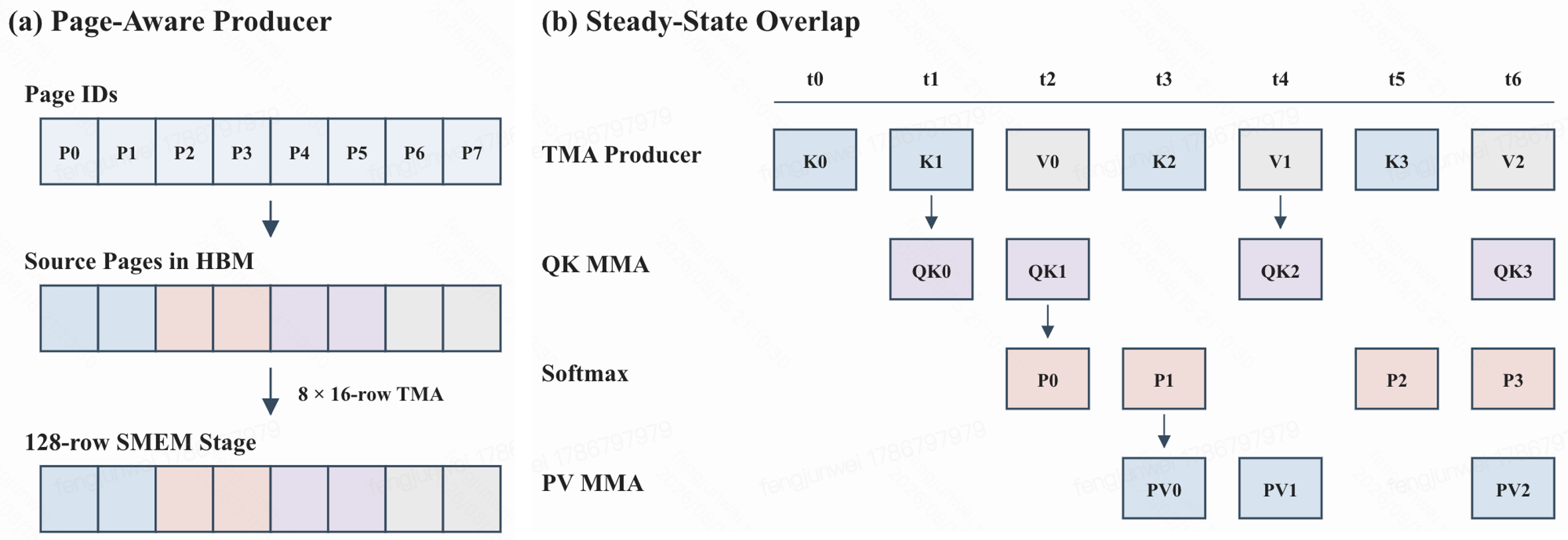}
    \caption{
    Page-guided K/V loading. Page16 regions are gathered from the
    original K/V tensors directly into the shared-memory tile consumed
    by the attention kernel. No compacted K/V tensor is materialized
    in HBM.
    }
    \label{fig:tma-pipeline}
\end{figure}

Each entry in the page table identifies a contiguous 16-row region in
the original K/V tensors. An $N128$ stage is assembled from eight such
regions directly in shared memory. The Q64 path uses two adjacent
$N128$ page groups to construct its $N256$ stage.

The shared-memory representation seen by the consumer is unchanged.
After the producer has assembled a stage, the existing QK matrix
multiply--accumulate (MMA), online-softmax, and PV MMA pipeline operates on a regular large tile.
Page indirection is therefore confined to the producer side.

Fully occupied logical blocks also retain a regular access pattern.
Likewise, backward detects aligned eight-page blocks and uses the
original full-tile transfer path for them. Page-wise transfers are
required only when physical tiles contain pages from different or
partial logical blocks.

\subsubsection{Exact handling of partial pages}

A Page16 tail may contain fewer than 16 valid rows, and dummy pages
introduced by alignment contain no valid rows. Zero-filling these rows
is not sufficient for attention, because a zero key would still
contribute a finite value to the softmax normalization.

KPage therefore masks invalid scores before the row maximum. Let
$S_{ij}$ and $\widetilde S_{ij}$ be the attention scores before and
after masking, and let $v_j$ indicate whether physical K/V row $j$
is valid:

\begin{equation}
\widetilde S_{ij}=
\begin{cases}
S_{ij},  & v_j=1,\\
-\infty, & v_j=0,
\end{cases}
\label{eq:score_mask}
\end{equation}

This ordering is required for exact attention semantics. Since invalid
scores are removed before the maximum and exponential, they contribute
neither to the online-softmax maximum nor to its denominator.

KPage stores validity at Page16 granularity using compact bit masks.
Two Page16 masks are packed into one 32-bit word. The page-table builder
also records where the fully valid region ends, so validity metadata is
only accessed for physical tiles that overlap the partial suffix.

The distinction is reflected in the kernel implementation. For fully
valid tiles, KPage retains the fused score-loading and row-maximum
path used by the underlying kernel. For a masked tile, the scores are
loaded, the packed validity predicate is applied, and the row maximum
is then computed from the masked values. The additional masking work
is therefore restricted to tiles that actually contain partial or
dummy rows.

For partial K/V pages, the unused suffix is also zeroed before the
attention launch so that the Page16 TMA transfer remains a regular
contiguous load. This zeroing is an implementation convenience; exact
softmax semantics are provided by Eq.~\eqref{eq:score_mask}.

\subsection{K/V-Owned Paged Backward}
\label{sec:backward}

Forward page packing cannot be applied to backward without changing
the work organization. In forward, a Q block owns one sparse walk and
only reads K/V. In backward, several Q blocks can contribute to the
same dK and dV rows. KPage therefore constructs a separate physical
layout centered on K/V tiles.

\begin{figure}[htbp]
    \centering
    \includegraphics[width=\linewidth]{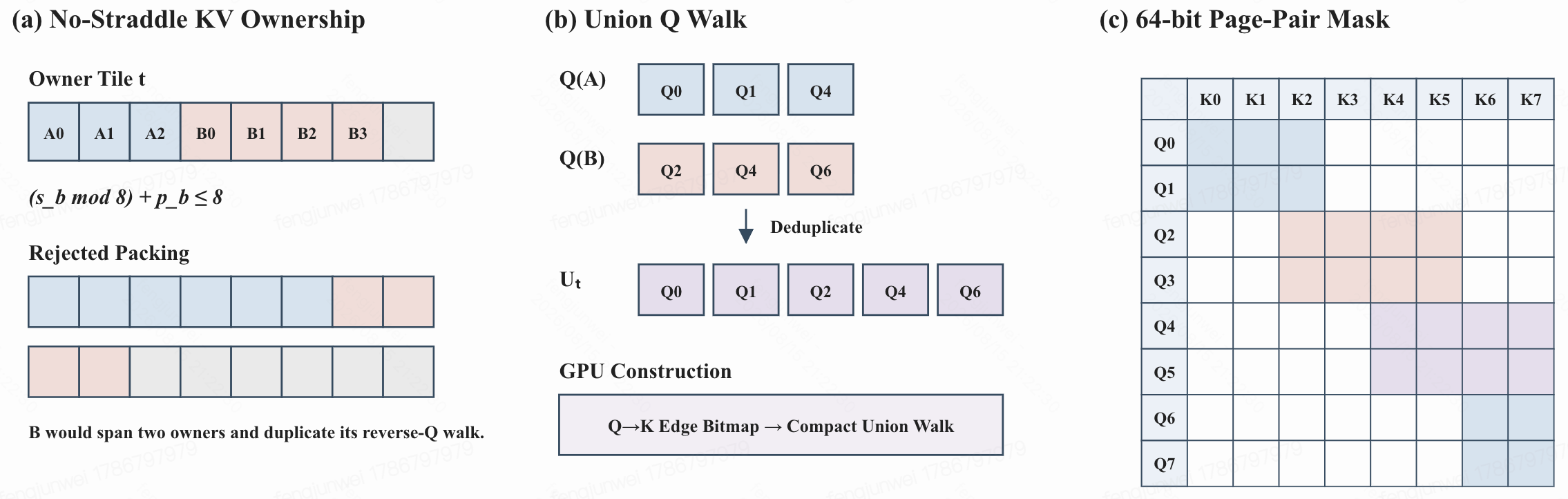}
    \caption{
    K/V-owned paged backward. K/V Page16 regions are packed into
    128-row owner tiles. Each owner traverses the Q pages connected
    to its constituent logical K/V blocks. Page-pair metadata restores
    the original sparse adjacency after packing.
    }
    \label{fig:paged-backward}
\end{figure}

\subsubsection{K/V owner packing}

A physical K/V owner contains eight Page16 slots. KPage allows pages
from multiple partial logical blocks to share an owner, but does not allow one logical block to cross an owner boundary. If $s_b$
is the starting page-slot index of block $b$, the layout requires

\begin{equation}
(s_b\bmod8)+p_b\le8 .
\label{eq:no_straddle}
\end{equation}

The constraint keeps all pages of one logical K/V block associated with
the same physical owner. Its reverse-Q neighborhood therefore needs to
be constructed only once for that owner.

Fully occupied blocks already contain eight pages and are kept aligned
to a complete owner tile. Partial blocks are packed into the remaining
owner tiles. This also provides a full-tile fast path: when the eight
page identifiers correspond to one aligned logical block, the backward
kernel uses the regular K/V tile load and store path rather than eight
independent page transfers.

\subsubsection{Reverse-Q construction and Q packing}

A packed owner can contain pages from several logical K/V blocks.
Backward must visit every Q block connected to any of these blocks.

Let $\mathcal{B}_t$ be the logical K/V blocks packed into owner $t$,
$\mathcal{Q}(b)$ the Q blocks connected to block $b$, and
$\mathcal{G}_g$ a group of Q blocks. KPage visits their reverse
neighborhood

\begin{equation}
\mathcal{U}_{t,g}
=
\mathcal{G}_g
\cap
\bigcup_{b\in\mathcal{B}_t}\mathcal{Q}(b).
\label{eq:reverse_union}
\end{equation}

The grouping bounds the amount of serial Q work assigned to one
backward work item for long sequences.

The reverse relation is constructed from the original Q-to-K sparse
edges. KPage scatters these edges into an owner-indexed bitmap and then
compacts the nonempty Q entries. This avoids constructing a separate
K-to-Q edge array followed by a global sort.

Q data are subsequently represented at Page16 granularity as well.
Full Q blocks remain aligned so they can use the regular 128-row load
path. Pages from partial Q blocks are packed into the remaining
physical Q steps. Thus, backward removes padding on both sides while
retaining the $128\times128$ MMA shape.

\subsubsection{Restoring the sparse graph after packing}

The reverse union introduces physical Q--K pairs that need not exist in
the original sparse graph. For example, an owner may contain pages from
two logical K/V blocks while a Q block is connected to only one of
them. Executing the full Cartesian product of the packed pages would
therefore introduce incorrect attention edges.

For a physical backward step, let $q_j$ and $k_i$ denote the logical
blocks owning Q-page $j$ and K/V-page $i$, and let $\mathcal{E}$ be
the original sparse block graph. The page-pair adjacency is

\begin{equation}
A_{ji}
=
\mathbf{1}\!\left[(q_j,k_i)\in\mathcal{E}\right],
\label{eq:page_adjacency}
\end{equation}

where $\mathbf{1}[\cdot]$ is the indicator function.

A physical step contains eight Q pages and eight K/V pages, so the
complete $8\times8$ adjacency matrix requires 64 bits and is stored in
one machine word. The kernel reads this word and masks page pairs with
$A_{ji}=0$ before reconstructing the attention probability. These
pairs therefore produce zero probability and zero score gradient.

When all 64 page pairs are valid, the mask is represented by an
all-valid value and the additional masking path is skipped. Packing
therefore does not impose page-pair tests on regular full-tile cases.

The resulting dQ, dK, and dV fragments are written back through the
existing FP32 accumulation and postprocessing layout. For mixed owner
tiles, KPage scatters results by Page16 identifier; for aligned full
tiles, it retains the regular bulk path.

\subsection{Metadata Construction and Kernel Integration}
\label{sec:implementation}

The forward layout depends on the sparse graph and valid-row counts,
not on the values of $Q$, $K$, or $V$. The page table, Q64/Q128 route
assignment, full-tile boundary, and suffix validity masks can therefore
be reused while the sparse layout is unchanged.

Forward metadata construction scans the selected sparse edges to count
the required Page16 regions, partitions active Q blocks into the two
execution routes, and emits the page and validity tables. The generated
metadata is consumed directly by the persistent forward kernels; no
runtime K/V compaction buffer is allocated.

Backward metadata is constructed lazily because it is unnecessary for
inference. The backward builder packs K/V owners, derives their
reverse-Q neighborhoods, packs the corresponding Q pages, and emits
the page-pair adjacency masks. The resulting layout can likewise be
reused while the sparse graph and block occupancies remain unchanged.

KPage retains the large-tile arithmetic of the reference NVIDIA BSA
implementation. Forward changes the K/V producer, introduces
Q64/Q128 routing, and adds validity handling for the partial page
suffix. Backward changes the physical ownership and traversal while
retaining the $128\times128$ attention-gradient computation. The
sparse-selection algorithm and attention equations are unchanged.

\subsubsection{Independent Q/K partitions}
\label{sec:independent_partitions}

The shared blk128 layout used by the representative workloads is a special
case of the page representation.  More generally, for each head let
$c_q^Q\in[0,128]$ denote the valid rows of logical Q block $q$, and
let $(c_k^K,o_k^K)$ denote the valid-row count and physical row offset
of logical K/V block $k$.  We require $o_k^K$ to be Page16 aligned and
the page footprints of distinct K/V blocks to be disjoint.  The sparse
graph $\mathcal{K}_q$ indexes this K/V partition; the numbers and
boundaries of Q and K/V blocks need not match.  Clusters larger than
128 rows are represented by multiple logical blocks with the same
sparse neighborhood.

Forward routing depends only on $c_q^Q$, while the K/V walk emits page
identifiers starting at $o_k^K/16$ and masks the final page according
to $c_k^K$.  Backward independently packs the K/V partition and builds
reverse-Q neighborhoods over the $n_Q\times n_{KV}$ sparse graph,
where $n_Q$ and $n_{KV}$ count the logical blocks in each partition.
Padding-state cleanup likewise uses Q sizes for dQ and K/V sizes for
dK and dV.  The attention kernels consume the same page identifiers,
validity masks, and page-pair adjacency words as in the shared case.
The shared layout is recovered with
$n_Q=n_{KV}$, $c_b^Q=c_b^K$, and $o_b^K=128b$.

\begin{algorithm}[t]
\caption{KPage Execution}
\label{alg:kpage}
\small
\begin{algorithmic}[1]

\Require Sparse graph $\{\mathcal{K}_q\}$,
         Q sizes $\{c_q^Q\}$, K/V sizes and offsets
         $\{c_k^K,o_k^K\}$,
         and $Q,K,V$

\Statex
\Statex \textbf{Forward}

\For{each query block $q$}
    \State form a K/V page walk from
           $\{(c_k^K,o_k^K):k\in\mathcal{K}_q\}$
    \State place full Page16 regions before partial tail pages
    \State choose the Q64 or Q128 execution path from $c_q^Q$
\EndFor

\For{each active query block $q$}
    \State gather its K/V pages directly into shared memory
    \State mask invalid rows in partial or alignment pages
    \State execute large-tile attention
\EndFor

\Statex
\Statex \textbf{Backward}

\State pack K/V pages into non-straddling physical owner tiles

\For{each K/V owner}
    \State construct the union of its reverse-Q neighborhoods
    \State pack the corresponding Q blocks at Page16 granularity
    \State record the original Q--K adjacency for each physical step
\EndFor

\For{each K/V owner and Q-group}
    \State gather the packed Q and K/V pages
    \State remove page pairs absent from the original sparse graph
    \State execute the $128\times128$ attention-gradient computation
\EndFor

\end{algorithmic}
\end{algorithm}


\section{Experimental Results}
\label{sec:evaluation}

\subsection{Experimental Setup}
\label{sec:eval_setup}

We evaluate KPage using a controlled padding comparison and five
representative sparse-attention workload cases covering varied sequence
lengths, sparsity levels, and block-occupancy patterns.

\paragraph{Controlled workloads.}
We construct two groups of matched-scale workloads corresponding to
5, 10, 15, and 30\,s video sequences.

The first group contains no padding: all selected logical blocks are
fully occupied. The second group uses three-dimensional layouts that
produce partially occupied boundary blocks. Across these workloads,
padding accounts for 20.8--22.8\% of the physical sequence, and
approximately one third of the logical blocks are partial.

Both groups use the same sparse-selection procedure and comparable
sequence scales. Within each workload, KPage and NVIDIA BSA consume
the same sparse block graph and the same valid-row metadata.

The two groups are not identical sparse graphs because the underlying
layouts produce different numbers of logical blocks under the same
selection procedure. We therefore use this experiment to compare two
execution regimes: fully occupied blocks and layouts with substantial
partial-block fragmentation.

\paragraph{Representative workload cases.}
We additionally evaluate five sparse-attention workload cases spanning
different sequence scales and sparsity regimes. Their physical sequence
lengths range from 470K to 629K tokens, and their retained valid
sparsity patterns vary across the cases, as do the proportions of
partially occupied blocks. These cases evaluate page compaction
across different sparse-attention settings.

The controlled and representative workloads use eight attention heads;
the SVG2 traces use the model's heads (12 for Wan 1.3B, 40 for
Wan 14B).  All workloads use head dimension 128 and BF16 inputs.

\paragraph{Baselines and software.}
FA4 serves as the dense-attention reference in Figures~\ref{fig:kpage_forward}
and~\ref{fig:bwd-speedup}, while NVIDIA BSA serves as the principal
block-sparse baseline. The FA4-BSA column in Table~\ref{tab:generality_performance}
is a sparse kernel and must not be confused with dense FA4. For every
reported forward comparison, KPage and BSA receive
identical sparse metadata and valid-row information.
Section~\ref{sec:eval_generality} additionally evaluates layouts
reconstructed from open-source SSTA, STA, and VSA systems.  It compares
KPage against NVIDIA BSA, FastVideo's native and Triton VSA kernels,
upstream FA4 block sparsity, and FlexAttention under the
same GPU execution environment.  It also evaluates the independent-partition
extension on a trace from the official SVG2 Wan 14B/720p pipeline,
including SVG2's released variable-block kernel and the FA4-BSA path.
The configuration matches SVG2's published experiment, allowing
its reported FA3-based results to be contrasted with the FA4-based
measurements in our setup. These comparisons refer to different
execution environments.

Experiments are conducted on a single GPU using CUDA 13.2. KPage is implemented with CUTLASS DSL, and NVIDIA BSA is
provided by cuDNN Frontend 1.28.0. Since the generated BSA kernel can
depend on the CUTLASS DSL version, we benchmark compatible BSA
configurations using CUTLASS DSL 4.5.0 and 4.7.1 and report the faster
result for each workload.  The open-source layouts in
Section~\ref{sec:eval_generality} run KPage with CUTLASS DSL 4.4.2 and
BSA and upstream FA4 (4.0.0b33) with CUTLASS DSL 4.7.1.

\paragraph{Timing methodology.}
Unless otherwise stated, we report bare attention-kernel execution
time. Sparse-graph construction, KPage page-table construction,
Q64/Q128 route assignment, and backward layout construction are
performed outside the timed region because these structures depend
only on the sparse graph and block occupancies and can be reused while
the layout remains unchanged.

The timed region contains the complete attention kernel, including
page-table traversal, page-guided K/V loading, partial-page masking,
online softmax, and output computation. No compacted K/V tensor is
materialized in HBM.

For the controlled workloads in
Table~\ref{tab:padding_ablation}, we report the median of 20 timed
iterations after 5 warmup iterations. For the five representative workload cases,
we report the median of 10 timed iterations after 3 warmup iterations.
For the SVG2 study, we report the median of 20 timed iterations after
3 warmup iterations.
All measurements use synchronized wall-clock timing on a single GPU
and a single execution stream.

\subsection{Numerical Correctness}
\label{sec:eval_correctness}

We validate KPage against an exact FP32 masked-attention reference.
For each workload, we sample query rows from multiple heads and logical
blocks and compare both forward outputs and backward gradients using the
same sparse graph and valid-row metadata.

Across the evaluated workloads, KPage forward output errors remain in
the same range as the BF16 BSA baseline, with maximum absolute errors
on the order of $10^{-4}$. The log-sum-exp (LSE) error remains below
$2\times10^{-6}$. For backward, the cosine similarity of
$dQ$, $dK$, and $dV$ is at least 0.99999 across the sampled cases.

These results confirm that page compaction changes only physical data
placement and execution order; invalid rows and page pairs are excluded
before softmax or probability reconstruction, preserving the original
sparse-attention semantics.

For the independent SVG2 partitions, sampled forward outputs over all
heads have relative $\ell_2$ error $2.34\times10^{-3}$ (Wan 1.3B
traces) and $2.35\times10^{-3}$ (Wan 14B trace) against the FP32
masked reference, comparable to the released variable-block kernel's
$2.27\times10^{-3}$ and $2.30\times10^{-3}$.  A full-row backward
check on one head of a Wan 1.3B trace gives relative $\ell_2$ errors
of $(2.37,2.35,2.32)\times10^{-3}$ for dQ, dK, and dV, respectively,
against $1.66\times10^{-3}$ for BF16 scaled dot-product attention
(SDPA) on the same inputs.
\subsection{Forward Performance}
\label{sec:eval_forward}

Table~\ref{tab:padding_ablation} compares KPage and NVIDIA BSA under
fully occupied and 3D-partial layouts.

On the padding-free layouts, all logical blocks are full. The
per-workload BSA/KPage speedups are $1.11\times$, $1.03\times$,
$1.06\times$, and $0.94\times$, giving a geometric mean of
$1.03\times$. KPage and BSA are therefore effectively on par when
page compaction is unnecessary.

This comparison also rules out a general kernel-speed advantage as the
source of the padded-layout result. On these padding-free workloads,
KPage's pipeline alignment schedules approximately 1.4--1.5\% more
128-row tile equivalents than BSA, while the measured performance
remains close to parity.

Under the $2\times8\times8$ 3D layout, boundary cubes introduce
20.8--22.8\% padding. KPage achieves $1.73\times$, $2.20\times$,
$2.44\times$, and $2.95\times$ speedup over BSA on the four
workloads, respectively, with a geometric mean of $2.29\times$.

The two layout families do not contain identical sparse graphs because
their different block constructions produce different numbers of
logical blocks under the fixed top-64 selector. The controlled result
should therefore be interpreted as a comparison between two execution
regimes: KPage is near parity with BSA when logical blocks are fully
occupied, but provides a substantially larger advantage when the same
selection procedure produces partial blocks under 3D boundary
fragmentation.

We observe the same behavior across the five representative workload
cases. KPage achieves per-case bare forward-kernel speedups of
$1.90\times$, $1.35\times$, $1.94\times$, $2.06\times$, and
$1.68\times$ over NVIDIA BSA, with a geometric mean of
$1.77\times$.

\begin{figure}[htbp]
    \centering
    \includegraphics[width=\linewidth]{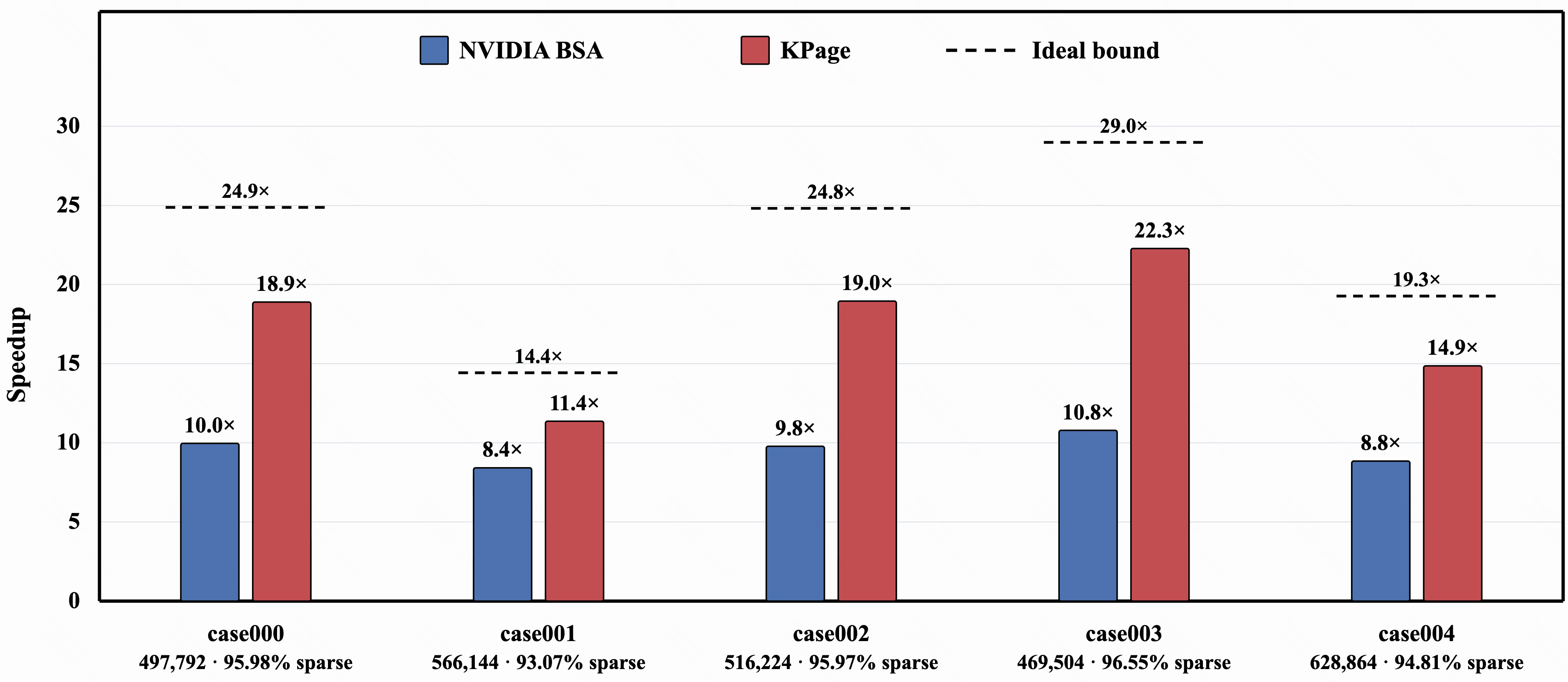}
    \caption{
    Forward bare-kernel speedup relative to dense FA4 across five
    representative workload cases. Dashed lines denote the ideal bound
    from the valid-pair sparsity shown for each case. KPage reaches
    approximately 76--79\% of this bound.
    }
    \label{fig:kpage_forward}
\end{figure}

Figure~\ref{fig:kpage_forward} compares KPage with the ideal speedup
derived from each case's valid-pair sparsity. Based on the plotted
values, KPage reaches approximately 76--79\% of this ideal sparse-work
bound, leaving approximately $1.26\times$--$1.32\times$ headroom.

\subsection{Reduction of Padding-Induced Physical Work}
\label{sec:eval_work}

We reconstruct the forward execution geometry directly from the
complete q2k metadata and valid-row counts of the five representative
workload cases. The useful sparse computation contains 160.611G valid
query--key token pairs.

If every selected logical blk128 pair is represented by a full
$128\times128$ physical tile, the five workloads expose 334.109G
physical query--key positions, corresponding to 48.1\% useful-work
utilization.

Page16 K/V compaction provides the largest reduction. Keeping the
query side at $M128$ while compacting selected K/V blocks at Page16
granularity reduces the physical pair capacity from 334.109G to
199.689G.

KPage additionally applies Q64/Q128 routing. Query blocks with at most
64 valid rows use the $M64\times N256$ path instead of a full $M128$
query tile. This further reduces the physical pair capacity from
199.689G to 177.548G, an additional 11.1\% reduction relative to the
M128-only paged layout.

\begin{figure}[htbp]
    \centering
    \includegraphics[width=\linewidth]{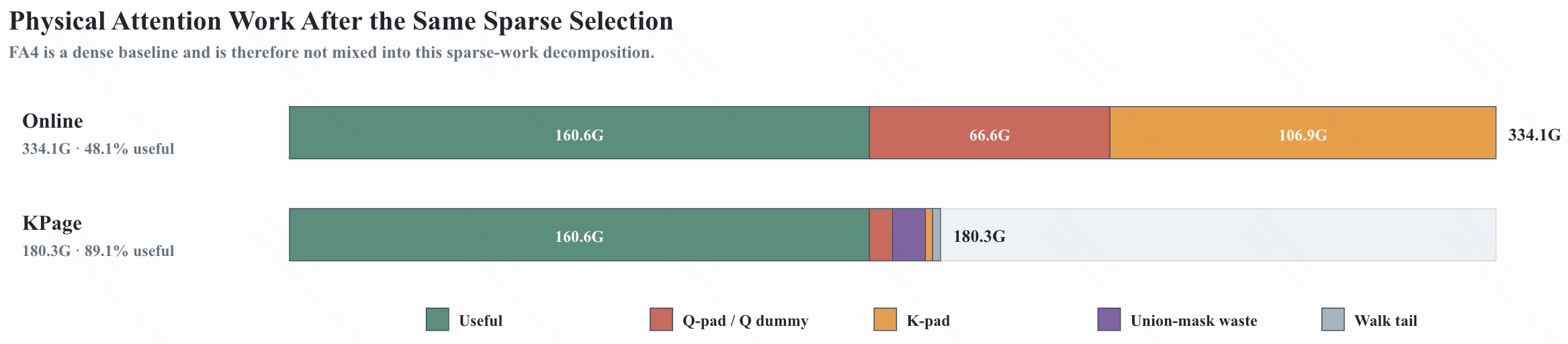}
\caption{
Physical attention-work breakdown across five representative
workload cases, preserving 160.6G useful pairs. The illustrated
whole-block and packed totals are 334.1G and 180.3G (48.1\% and
89.1\% useful), respectively; the latter includes the displayed
union-mask and walk-tail terms. The separate nominal tile-capacity
accounting in the text gives 177.5G (90.5\% useful).
}
    \label{fig:physical-work}
\end{figure}

Overall, this nominal packed-tile accounting reduces physical pair
capacity by 46.9\% relative to the whole-block layout and raises
pair-capacity efficiency from 48.1\% to 90.5\%, while preserving
the same 160.611G useful sparse interactions. The companion breakdown
in Figure~\ref{fig:physical-work} also depicts residual union-mask
and walk-tail overhead, yielding a slightly higher 180.3G total
(89.1\% useful) under that overhead-inclusive accounting.

These quantities characterize the physical capacity implied by the
corresponding execution layouts. In particular, the 334.109G
whole-block value should not be interpreted as a direct measurement of
NVIDIA BSA forward FLOPs: BSA forward supports valid block-size
metadata and preserves exact partial-block semantics. The direct BSA
comparison is therefore given by the measured wall-clock results in
Section~\ref{sec:eval_forward}.

The nominal accounting separates the contributions of the two
forward mechanisms. Most of the reduction comes from Page16 K/V
compaction, while Q64/Q128 routing removes a further 11.1\% of
nominal physical pair capacity by reducing Q-side padding. Figure~\ref{fig:physical-work}
separately visualizes the additional overhead categories; its 180.3G
total is therefore not the same accounting quantity as 177.5G.


\subsection{Backward Performance}
\label{sec:eval_backward}

For training-oriented execution, we additionally measure the combined
forward-and-backward bare-kernel time.

The padding-free workloads provide an exact comparison because every
logical block is fully occupied and NVIDIA BSA backward does not require
partial-block semantics. The per-workload forward--backward BSA/KPage
ratios are $0.98\times$, $0.98\times$, $1.00\times$, and
$0.98\times$, giving a geometric mean of $0.99\times$. KPage and BSA
are therefore effectively identical when no partial blocks are present.

Under the 3D-padded layouts, the corresponding timing ratios are
$1.36\times$, $1.33\times$, $1.36\times$, and $1.44\times$,
with a geometric mean of $1.37\times$.

These 3D-padded backward numbers are reported only as a hardware timing
reference. The current NVIDIA blk128 BSA backward path does not support
variable \texttt{block\_sizes} and therefore executes whole-block
semantics on partial layouts. They should not be interpreted as an
exact numerical comparison of the two backward operators.

Our primary quantitative comparison against NVIDIA BSA is therefore
the forward result, where both implementations support the same
valid-row semantics and execute the same sparse graph.

\begin{figure}[htbp]
    \centering
    \includegraphics[width=\linewidth]{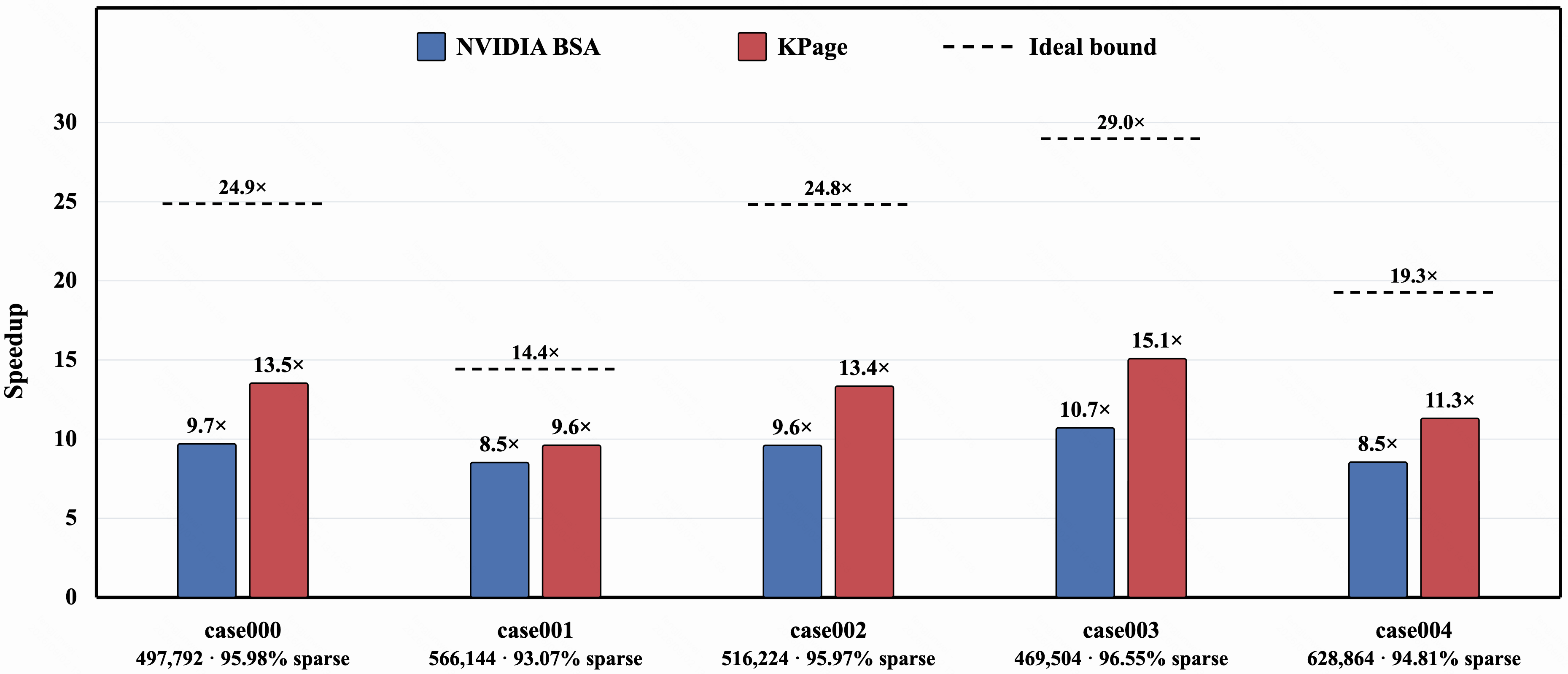}
    \caption{
    Forward-plus-backward kernel speedup relative to dense FA4 on the
    five representative workload cases (case000--case004), with the
    corresponding valid-pair-density ideal bound. This plot is distinct
    from the controlled 5--30\,s padding comparison. NVIDIA BSA's
    backward result is a hardware timing reference where partial-block
    semantics differ; it is not an exact numerical comparison.
    }
    \label{fig:bwd-speedup}
\end{figure}

Figure~\ref{fig:bwd-speedup} separately reports the five representative
cases, normalizing forward-plus-backward kernel timings to dense FA4.
These are distinct from the four controlled padding-length comparisons
above; the BSA bars are timing references where partial-block backward
semantics are not identical.

\subsection{Generality Across Open-Source Sparse Layouts and Kernels}
\label{sec:eval_generality}

The preceding experiments cover several workload configurations. To
test whether KPage generalizes to other video models and sparse
selectors, we evaluate ten configurations from four open-source
systems.  The HunyuanVideo cases comprise SSTA at 480p and 720p with
121 or 241 frames, and STA at 720p with 117 frames
~\cite{wu2025hunyuanvideo15,zhang2025sta}.  The Wan cases comprise VSA
at 480p and 720p with $(4,4,4)$ or $(4,8,8)$ spatiotemporal cubes
~\cite{zhang2025vsa}, plus SVG2 at 480p on Wan 1.3B and at 720p on Wan
14B~\cite{yang2025sparsevideogen2}.  SSTA combines a 3D window with
top-$k$ selection, STA uses head-specific windows, VSA routes learned
3D cubes, and SVG2 independently clusters and permutes Q and K.

For SSTA, STA, and VSA, block occupancy follows the released geometry,
padding rule, resolution, and frame count; we synthesize Q, K, and V
because these quantities do not depend on sample content.  For SVG2,
we capture cluster sizes and selected cluster pairs from official
end-to-end inference runs.  Every case is converted to the exact
block-index and valid-row metadata consumed by KPage; no kernel or
schedule is specialized for a selector.

We compare with NVIDIA BSA~\cite{nvidia2026bsa}, FastVideo's native
CUDA and Triton VSA kernels~\cite{zhang2025vsa}, upstream FA4 block
sparsity~\cite{zadouri2026flashattention4},
FlexAttention~\cite{dong2025flexattention}, and SVG2's released
FlashInfer variable-block kernel (VBS)~\cite{ye2025flashinfer}.  On
the 128-row-aligned SSTA, STA, and VSA layouts, BSA, VSA-CUDA, FA4,
and FlexAttention run with 128-token blocks and VSA-Triton with
64-token blocks.  On the SVG2 layouts, whose clusters are far smaller
than 128 rows, each baseline uses its fastest exact configuration:
64-token blocks for VSA-CUDA, FlexAttention, and Triton, and 128-token
blocks for BSA and FA4, which offer no smaller exact block.  All runs
use BF16, head dimension 128, and the same GPU.
Metadata construction is excluded; each latency is the median of ten
iterations after three warmups (twenty iterations for the SVG2 rows).

In Table~\ref{tab:generality_performance}, a VSA tuple $(t,h,w)$ is
the temporal, latent-height, and latent-width cube.  \emph{Pad} is the
invalid-row fraction after blk128 placement, and \emph{Partial} is the
fraction of nonempty blk128 blocks that are not full.  For SVG2,
which has independent partitions, Pad reports Q/K padding and Partial
is marked \emph{indep.}

\begin{table*}[!t]
\centering
\caption{Fragmentation statistics and attention latency on layouts
from open-source systems.  Pad rows and Partial blocks are defined in
the text.  Best/KPage is the minimum latency among exact sparse
baselines divided by KPage latency; values above one indicate a KPage
speedup. Displayed latencies and speedups are rounded independently.  $^\dagger$BSA backward executes whole blocks and is reported
as a hardware timing reference but excluded from Best/KPage.
$^\ddagger$SVG2 uses independent Q/K partitions; its Pad rows entry
reports the Q-side and K-side blk128 padding (Q/K), averaged over the
traces of the row.  The released workload is inference-only, so no
forward--backward result is reported.  SVG2's released kernel is not a
column: it runs 4.43\,ms (1.3B) and 98.67\,ms (14B), see text.}
\label{tab:generality_performance}
\small
\renewcommand{\arraystretch}{1.08}

\textbf{(a) Forward time (ms)}\par
\vspace{2pt}
\setlength{\tabcolsep}{3pt}
\resizebox{0.86\textwidth}{!}{%
\begin{tabular}{@{}lrrrrrrrrr@{}}
\toprule
\textbf{Layout}
& \textbf{Pad rows}
& \textbf{Partial blocks}
& \textbf{NV BSA}
& \textbf{VSA-CUDA}
& \textbf{FA4-BSA}
& \textbf{Flex}
& \textbf{VSA-Triton}
& \textbf{KPage}
& \textbf{Best/KPage} \\
\midrule
SSTA-Hunyuan 480p/121f & 10.0\% &  18.3\% &  6.77 &  6.71 &  13.66 &  16.66 &  14.77 & \textbf{ 4.58} & \textbf{1.46$\times$} \\
SSTA-Hunyuan 480p/241f &  8.5\% &  15.5\% & 21.45 & 22.36 &  40.34 &  50.02 &  45.74 & \textbf{17.26} & \textbf{1.24$\times$} \\
SSTA-Hunyuan 720p/121f &  4.2\% &  12.2\% & 16.02 & 16.20 &  26.41 &  35.21 &  34.63 & \textbf{13.15} & \textbf{1.22$\times$} \\
SSTA-Hunyuan 720p/241f &  2.5\% &   9.1\% & 51.90 & 51.92 &  73.58 & 103.72 & 108.24 & \textbf{42.70} & \textbf{1.22$\times$} \\
STA-Hunyuan 720p/117f  &  0.7\% &   5.9\% & 56.03 & 56.02 &  75.52 & 114.78 & 126.39 & \textbf{49.72} & \textbf{1.13$\times$} \\
\addlinespace[2pt]
VSA-Wan 480p $(4,4,4)$ & 59.0\% & 100.0\% &  4.17 &  3.68 &  24.98 &  16.86 &   4.60 & \textbf{ 1.13} & \textbf{3.27$\times$} \\
VSA-Wan 720p $(4,8,8)$ & 10.5\% &  16.7\% & 14.77 & 15.00 &  30.60 &  34.91 &  30.13 & \textbf{11.90} & \textbf{1.24$\times$} \\
VSA-Wan 720p $(4,4,4)$ & 59.0\% & 100.0\% & 74.96 & 76.71 & 432.63 & 301.51 &  77.28 & \textbf{17.63} & \textbf{4.25$\times$} \\
\addlinespace[2pt]
SVG2-Wan 1.3B/480p$^\ddagger$ & 31.9/74.4\% & indep. & 5.63 & 2.76 & 35.88 & 14.48 & 5.21 & \textbf{1.43} & \textbf{1.93$\times$} \\
SVG2-Wan 14B/720p$^\ddagger$ & 7.8/30.7\% & indep. & 48.59 & 46.18 & 197.22 & 180.44 & 81.80 & \textbf{30.92} & \textbf{1.49$\times$} \\
\bottomrule
\end{tabular}
}

\vspace{6pt}
\textbf{(b) Forward + backward time (ms)}\par
\vspace{2pt}
\resizebox{0.86\textwidth}{!}{%
\begin{tabular}{@{}lrrrrrrrrr@{}}
\toprule
\textbf{Layout}
& \textbf{Pad rows}
& \textbf{Partial blocks}
& \textbf{NV BSA$^\dagger$}
& \textbf{VSA-CUDA}
& \textbf{FA4-BSA}
& \textbf{Flex}
& \textbf{VSA-Triton}
& \textbf{KPage}
& \textbf{Best/KPage} \\
\midrule
SSTA-Hunyuan 480p/121f & 10.0\% &  18.3\% &  24.99 &  25.71 &  30.55 &  55.68 &  62.90 & \textbf{22.30} & \textbf{1.15$\times$} \\
SSTA-Hunyuan 480p/241f &  8.5\% &  15.5\% &  79.44 &  84.50 &  99.96 & 168.38 & 194.23 & \textbf{73.51} & \textbf{1.15$\times$} \\
SSTA-Hunyuan 720p/121f &  4.2\% &  12.2\% &  58.30 &  61.83 &  69.54 & 118.31 & 148.21 & \textbf{56.46} & \textbf{1.10$\times$} \\
SSTA-Hunyuan 720p/241f &  2.5\% &   9.1\% & 181.52 & 190.97 & 220.97 & 352.78 & 462.62 & \textbf{176.49} & \textbf{1.08$\times$} \\
STA-Hunyuan 720p/117f  &  0.7\% &   5.9\% & 200.78 & 210.77 & 223.90 & 391.13 & 532.88 & \textbf{206.48} & \textbf{1.02$\times$} \\
\addlinespace[2pt]
VSA-Wan 480p $(4,4,4)$ & 59.0\% & 100.0\% &  13.80 &  13.62 &  34.57 &  51.41 &  19.16 & \textbf{ 5.99} & \textbf{2.27$\times$} \\
VSA-Wan 720p $(4,8,8)$ & 10.5\% &  16.7\% &  55.53 &  57.12 &  72.36 & 119.28 & 126.52 & \textbf{51.51} & \textbf{1.11$\times$} \\
VSA-Wan 720p $(4,4,4)$ & 59.0\% & 100.0\% & 239.90 & 269.03 & 633.66 & 908.29 & 331.75 & \textbf{99.10} & \textbf{2.71$\times$} \\
\bottomrule
\end{tabular}
}
\end{table*}

On the eight shared-partition layouts in
Table~\ref{tab:generality_performance}, padding ranges from 0.7\% to
59.0\% and the partial-block ratio from 5.9\% to 100\%.  KPage has the
lowest forward latency in all ten configurations.  Relative to the
fastest exact sparse baseline, the forward speedups are
$1.22$--$1.46\times$ for SSTA, $1.13\times$ for STA,
$1.24$--$4.25\times$ for VSA, and $1.49$--$1.93\times$ for SVG2.
For the eight training-capable SSTA, STA, and VSA cases, KPage also
has the lowest combined forward--backward time among exact baselines, with
$1.02$--$2.71\times$ speedup.  The nearly full-block STA case remains
close to parity, while the largest gains occur when every selected
block is partial.

SVG2 is the only special interface case.  Its independent Q/K
clusters require separate per-head Q sizes and K/V sizes and offsets;
KPage places the K/V clusters on Page16 boundaries without changing
the page kernels.  The 1.3B/480p row aggregates 15 middle-layer traces
with $C_q=300$ and $C_k=1000$.  The 14B/720p row matches the paper
configuration ($C_q=100$, $C_k=500$, 30\% dense warmup) on a released
prompt; its 31.9\% density is close to the reported 29.5--30.1\%.
The released SVG2 VBS interface exposes FA2/FA3 paths but does not
provide a matching device-specific backend for our execution environment.
It therefore uses FA2 and is slower than dense FA4 in our comparison.
We report VBS alongside the compatible sparse kernels available in
the same environment.  SVG2 reports $2.98\times$ H100
attention speedup over FA3 and $1.60\times$ end-to-end speedup
($1.89\times$ for SVG2-Turbo); our table reports attention-kernel
latency and makes no end-to-end KPage claim.

All exact forward kernels match an FP32 masked-attention oracle to
BF16 accuracy; training-capable kernels have gradient cosine
similarity of at least 0.9999.  The consistent ranking across these
models, resolutions, selectors, and block geometries shows that the
gain is not confined to the preceding workload configurations.

\subsection{Summary}

The controlled padding comparison shows that KPage remains near
parity with NVIDIA BSA when every logical block is full, with a
$1.03\times$ geometric-mean forward speedup. Under 3D-blocked
workloads, where boundary cubes introduce 20.8--22.8\% padding,
the geometric-mean forward speedup is $2.29\times$.

Across five representative workload cases, KPage achieves a
$1.77\times$ geometric-mean bare forward-kernel speedup over NVIDIA
BSA. Metadata-based execution accounting shows that Page16 K/V
compaction reduces physical pair capacity from 334.1G to 199.7G
pairs, while Q64/Q128 routing further reduces it to 177.5G. The
resulting pair-capacity efficiency increases from 48.1\% to 90.5\%
while preserving 160.6G useful sparse interactions.

These results show that KPage remains competitive when logical blocks
are already fully occupied and substantially reduces the cost of
partial-block fragmentation in three-dimensional video layouts.
The same behavior holds on layouts reproduced from open-source systems
(Section~\ref{sec:eval_generality}): against the fastest exact
block-sparse kernel available for each layout, KPage has the lowest
forward time in all eight SSTA, STA, and VSA cases.  The
geometric-mean forward ratios are $1.28\times$ for SSTA and
$2.58\times$ for VSA; a nearly full-block STA case remains close to
parity at $1.13\times$.  On the SVG2 Wan 1.3B middle-layer traces,
KPage is $1.93\times$ faster than the strongest exact sparse baseline.
On the paper-matched SVG2 Wan 14B/720p trace, KPage is
$1.49\times$ faster than the strongest exact sparse baseline and
$1.57\times$ faster than NVIDIA BSA. We do not infer a dense-FA4
speedup from the FA4-BSA column in Table~\ref{tab:generality_performance}.
\section{Conclusion}

KPage improves the execution of partially occupied block-sparse
attention without changing the sparse graph or shrinking the large
compute tile. It decomposes selected K/V blocks into Page16 regions,
packs valid pages across logical-block boundaries, and assembles the
resulting physical tiles directly during K/V loading. Query-adaptive
$M64\times N256$/$M128\times N128$ execution reduces Q-side padding,
while a K/V-owned paged layout extends the same compaction principle
to exact backward execution.

KPage is effectively on par with NVIDIA BSA when all logical blocks are
fully occupied ($1.03\times$ geometric mean), but reaches
$2.29\times$ geometric-mean forward speedup under 3D partial-block
fragmentation and $1.77\times$ across five representative workload cases.
Page16 K/V compaction and query routing reduce physical pair capacity
from 334.1G to 177.5G positions for 160.6G useful interactions,
increasing pair-capacity efficiency from 48.1\% to 90.5\%.
The open-source cross-system evaluation confirms that this behavior is
not pipeline-specific: across reconstructed SSTA, STA, and VSA layouts,
KPage achieves $1.13$--$4.25\times$ forward speedup over the strongest
exact baseline while remaining close to parity as blocks become full.
Separate Q/K partition metadata further extends the same page kernels
to SVG2's independent semantic clusters.  On the paper-matched
Wan 14B/720p trace, KPage is $1.49\times$ faster than the strongest
exact sparse baseline and $1.57\times$ faster than NVIDIA BSA.

The central result is that block-sparse attention need not choose
between fine-grained padding removal and large-tile execution:
fine-grained page packing can fill large physical tiles while preserving
the original sparse-attention semantics.
\bibliographystyle{ACM-Reference-Format}
\bibliography{kpage_refs}


\end{document}